\documentclass[12pt,nofootinbib,showpacs,prx,preprintnumbers,amsmath,amssymb,superscriptaddress,aps,longbibliography]{revtex4-2}
\usepackage{graphicx} 
\usepackage{booktabs}
\usepackage{siunitx}
\usepackage{makecell}
\usepackage{xcolor}   
\usepackage[colorlinks=true,citecolor=blue,urlcolor=blue,linkcolor=blue]{hyperref}
\usepackage[skip=0.1pt]{parskip} 
\usepackage{booktabs} 
\usepackage{orcidlink}
\usepackage{float} 
\usepackage{setspace}
\usepackage{textcomp}
\usepackage{amsmath}
\usepackage[utf8]{inputenc}
\usepackage{epstopdf}

\usepackage[margin=1in]{geometry}
\newcommand{\phase}[2]{\ensuremath{\mathrm{#1#2}}}

\usepackage{titlesec}

\titlespacing*{\subsubsection}{-1em}{1.5ex plus .2ex}{1ex}

\usepackage{amsmath}
\begin{document}

\title{First-Principles Investigation of the Al-V Phase Diagram}

\author{AKM Sadman Mahmud\orcidlink{0000-0002-1310-8622}}
\email{amahmud@andrew.cmu.edu}
\affiliation{Department of Physics, Carnegie Mellon University, Pittsburgh 15213, Pennsylvania, USA}

\author{Hassan Albuhairan}
\affiliation{Department of Physics, Carnegie Mellon University, Pittsburgh 15213, Pennsylvania, USA}

\author{Marek Mihalkovic}
\affiliation{Institute of Physics, Slovak Academy of Sciences, SK-84511 Bratislava, Slovakia}

\author{Michael Widom}
\affiliation{Department of Physics, Carnegie Mellon University, Pittsburgh 15213, Pennsylvania, USA}

\date{\today}

\begin{abstract}
The Al-V alloy system contains a number of phases including several with complex structures and at least two exhibiting sites of partial occupation. Through electronic density functional theory-based total energy calculations combined with methods of statistical mechanics, we examine the relative stability of phases at finite temperatures. We construct composition-continuous free energy models for the V-rich solid solution and for one of the complex intermetallic phases. In the V-rich region, we identify three ground states that transform to the solid solution at elevated temperatures. We also suggest that \phase{Al}{V_3} takes the Al15 structure as an intermediate-temperature phase stabilized by anharmonic vibrational free energy. 

\end{abstract}

\maketitle
\section*{Introduction}

Exploration of alloy phase diagrams can guide material discovery for targeting thermodynamic, electrical, and functional properties. Theoretical and computational modeling can explain the structures and thermodynamic stability of known phases and can potentially predict previously unknown structures. The reported phase diagrams of Al-V \cite{ASM_MPDS_AlV} contain unresolved inconsistencies that computational modeling can help to resolve. Many proposed diagrams ~\cite{carlson1955, pauling_file_ref_id_184268,  kaufman1989} imply a continuous low temperature stability of the V-rich solid solution in apparent violation of the third law of thermodynamics. Two of the line compounds, \phase{Al_{10}}{V} and \phase{Al_8}{V_5}, paradoxically exhibit partial or mixed occupation sites. One phase diagram~\cite{murray1989} includes a room temperature phase \phase{Al}{V_3}, whose existence remains experimentally uncertain~\cite{holleck1963kristallstruktur, kornilov1968compound, hartsough1971synthesis, leger1973pressure}. The phase boundary of the solid solution was estimated from simple thermodynamic modeling constrained by limited experimental data.

Some of the Al-V phases display interesting physical properties. \phase{Al_{10}}{V}, a cage-like structure with potential guest atoms, exhibits anomalous low-temperature behavior with specific heat and electrical resistivity following an Einstein-like temperature dependence \cite{CaplinNicholson1978JPhysF, CaplinGrunerDunlop1973PRL, albuhairan2025first}. \phase{Al}{V_3}, an A15 compound, is notable for potential superconductivity \cite{holleck1963kristallstruktur}. Al and V together can strengthen lightweight alloys such as Ti for aerospace applications.

\begin{figure}[H]
  \centering
  \includegraphics[width=\linewidth]{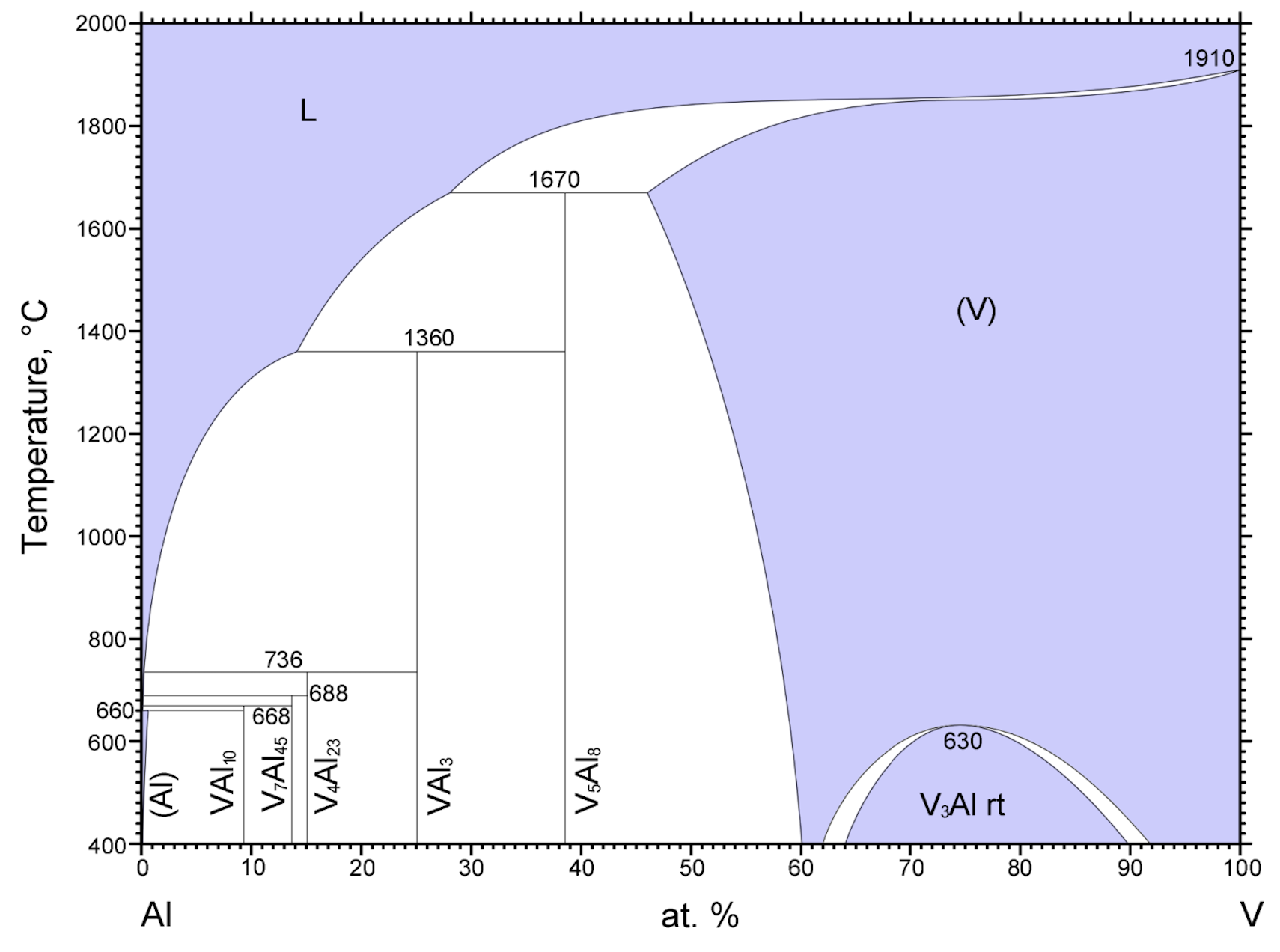}
  \caption{An experimentally informed Phase Diagram for Al-V \cite{murray1989}.}
  \label{fig:AlVPD}
\end{figure}

As assessed by Murray~\cite{murray1989} (see Figure~\ref{fig:AlVPD}), Al-V exhibits five stable Al-rich line compounds: complex cubic \phase{Al_{10}}{V} (also known as \phase{Al_{21}}{V_2} or \phase{Al_{11}}{V}) with Pearson types cF176 and cF184, monoclinic \phase{Al_{45}}{V_7} (Pearson mC104), hexagonal \phase{Al_{23}}{V_4} (Pearson hP54), tetragonal \phase{Al_3}{V} (Pearson tI8), and cubic $\gamma$ brass \phase{Al_8}{V_5} (Pearson cI52). Although Murray treated \phase{Al_8}{V_5} as a line compound ($x = 0.385$), she speculated about a range of homogeneity for \phase{Al_8}{V_5} at high temperatures, which was later experimentally determined~\cite{richter2000v}.  All the Al-rich structures other than \phase{Al_3}{V} possess Al- or V-centered icosahedral clusters, and \phase{Al_{10}}{V} also possesses a 16-atom Friauf polyhedron. The \phase{Al_{45}}{V_7}, \phase{Al_{23}}{V_4}, and \phase{Al_8}{V_5} structures can be considered as vacancy-ordered variants of BCC~\cite{lord2004gamma,Schubert,Leineweber}.

In the V-rich region of the phase diagram, a body-centered solid solution extends from approximately 50\% up to pure V. The putative A15 phase \phase{Al}{V_3} also lies in this V-rich region and is represented with a broad composition range, despite its Frank-Kasper nature that is better suited to a line compound. In addition, a tetragonal compound of the same composition has been reported experimentally at lower temperatures \cite{leger1973pressure}.

Our work seeks to understand the stability of known phases and potentially discover new stable structures. We combine first-principles density functional theory total energies with statistical mechanics tools to calculate free energy at finite temperature. We examine the relative stability between Al-rich alloys close in composition and also the free energy competition between these line compounds and the V-rich solid solution. For the latter, we introduce a composition- and temperature-dependent free energy model to determine its composition range. We resolve the apparent third law violation by predicting new V-rich line compounds which out-compete the solid solution at low temperatures.

\section*{Methods}

\subsection*{DFT}

We perform first-principles total-energy calculations using the Vienna Ab initio Simulation Package (VASP~\cite{KresseJoubert1999PRB, KresseFurthmueller1996PRB}) within the PBE generalized-gradient approximation~\cite{perdew1992atoms}.  Our Al pseudopotential is the standard choice with valence 3 while our V potential has valence 13, as it treats the 3s and 3p semicore levels as valence. We relax all atomic positions and lattice parameters using a fixed plane-wave cutoff energy of 340 eV in order to reduce the Pulay stress artifact. Reciprocal-space evaluation of the PAW projectors is used for reliable forces, and the calculations are run in the accurate precision mode to avoid wrap-around errors. We increase the $k$-point densities until the energies converge to within 0.1 meV/atom. Thermal expansion is not considered. 

We obtain most Al-rich structures from the ICSD Database \cite{ICSDWebsite}. We predict other possible ground state structures using the ATAT method \cite{van2002alloy, van2009multicomponent}. Details of these structures are given in Appendix~\ref{app:structures}. We test for potential magnetism using an initial magnetic moment of 3 \(\mu_\text{B}\) per V atom.  We obtain phonon vibrational modes via density functional perturbation theory (DFPT) applied to fully relaxed supercells containing at least 104 atoms. For the special case of \phase{AlV_3}{.cP8}, we compute vibrational free energies with anharmonic contributions using the temperature-dependent effective potential (TDEP) method \cite{Hellman2013,Knoop2024TDEP}.   


\subsection*{Free Energy Modeling}

Under condition pressure $P=0$, the Gibbs free energy \(G=H-TS\) and the Helmholtz free energy \(F=U-TS\) are equivalent. For any structure \(\tau\),  assuming the decoupling of electrons and phonons and taking into account the excitations of the electronic state, the free energy per atom can be summed as 
\begin{equation}
    F^{(\tau)} = \Delta H_{f}^{(\tau)}+F_{\text{vib}}^{(\tau)}(T)+F_{\text{elec}}^{(\tau)}(T).
\end{equation}
Here $\Delta H_{f}^{(\tau)}$, \(F_{\text{vib}}^{(\tau)}\), and \(F_{\text{elec}}^{(\tau)}\) are the enthalpy, vibrational, and electronic free energies per atom of the structure \(\tau\).

The 0 K enthalpy $\Delta H_{f}^{(\tau)}$ is calculated relative to a tie-line that connects the ground state configurations of the pure elements \phase{Al}{.cF4} and \phase{V}{.cI2}. For a composition of stoichiometry $\mathrm{Al}_{1-x}\mathrm{V}_x$,  we define
\begin{equation}
\Delta H_{f}(x) = H_f\!\left(\mathrm{Al}_{1-x}\mathrm{V}_x\right) - (1-x)\,H_f(\mathrm{Al}) - x\,H_f(\mathrm{V}).
\end{equation}
The convex hull of $\Delta H_{f}(x)$ determines the set of stable structures at low temperature. We also define $\Delta E^{(\tau)}$ as the enthalpy of structure $\tau$ relative to the convex hull at the same composition.

We calculate the vibrational free energy per atom \(F_{\mathrm{vib}}(T)\) based on the non-zero harmonic oscillator frequencies \(\omega_i\) of structure \(\tau\) with the total number of atoms of the supercell, \(N_{\text{scell}}\):
\begin{equation}
    F_{\mathrm{vib}}^{(\tau)}(T)=\frac{k_B T}{N_{\text{scell}}} \sum_{i=1}^{3N_{\text{scell}}-3}
\ln\!\left[2\sinh\!\left(\frac{\hbar\omega_i^{(\tau)}}{2k_B T}\right)\right].
\end{equation}
We calculate \(F_{elec}^{(\tau)}\) from the electronic density of states \(D(E)\) taking care to include the temperature-dependent electron chemical potential~\cite{widom2018modeling}. 

The configuration space of a solid solution can be partitioned into basins $\tau$. Each basin \(\tau\), associated with a distinct local minimum of the potential-energy surface, corresponds to a distinct atomic configuration with symmetry-equivalent multiplicity \(\Omega^{(\tau)}\). At sufficiently low temperatures, the system performs small oscillations within a basin. The partition function of the solid solution~\cite{widom2018modeling},
\begin{equation}
\label{eq:partition}
    Z_{\text sol} = \sum_\tau \Omega^{(\tau)} e^{-\beta NF^{(\tau)}} ,
\end{equation}
yields its free energy per atom
\begin{equation}
    F_{\text{sol}} = - \frac{k_BT}{N}  \ln{Z_{\text sol}}.
\end{equation}
We call this method full enumeration of configurations and note that it is similar to the approach known as zentropy~\cite{zentropy}.

For the bcc solid solution phases, we first consider the \(2\times2\times2\) superstructures of the 2-atom body-centered unit cell, resulting in a total of 16 atoms. This allows us to sample the solid solution in a variety of compositions, $x=n/16$: we choose $n=8,10-14,$ and $ 16$. Taking advantage of the mixed occupancy in all sites, we enumerate all possible symmetry-distinct configurations \(\tau\) in the ensemble at each composition using enumlib \cite{hart2012generating}. The total numbers of such configurations, with increasing values of \(n\) range from 59 to 1. We then create 128-atom \(2\times 2\times2\) supercells of these 16-atom cell configurations. For any configuration \(\tau\) with imaginary phonon frequencies, indicating dynamical instability, we approximate the corresponding \(F^{(\tau)}_{vib}\)
by using the average vibrational free energy of the dynamically stable configurations at the same composition.

\phase{Al_8}{V_5}.cI52 has four sublattices, two with mixed occupancies, in its 26-atom primitive cell. Utilizing these two partially occupied sublattices, we enumerate the symmetry-distinct configurations. The total number of symmetry-independent configurations is 18 in this case. We create \(2\times2\times2\) supercells of these primitive cells, resulting in a total of 208 atoms in the supercells for these configurations. We apply the same strategy for two neighboring cI52 compositions  \phase{Al_{15}V_{11}}{} (\(x=0.423\)) and \phase{Al_{17}V_{9}}{} (\(x=0.346\)).

The entropy per atom of a finite-size random mixture lies below its thermodynamic limit.
Consider $N$ sites with composition $x$ for atom type A and $1-x$ for atom type B. The number of possible configurations is
\begin{equation}
\Omega=\frac{N!}{(xN)!\,\bigl[(1-x)N\bigr]!}.
\end{equation}
Using the Stirling approximation to subleading order, \(\ln N! \simeq N\ln N - N + \frac12\ln(2\pi N),\)
one obtains the entropy per atom
\begin{equation}
S \simeq -
k_B \Big[x\ln x+(1-x)\ln(1-x)\Big]
-\frac{k_B}{2N}\ln\!\big[2\pi x(1-x)N\big].
\end{equation}
The final term reduces the entropy by 15\(\%\) for \(N=16\) at \(x=1/2\) and the fraction diverges as $x$ approaches 0 or 1.  This formula can be generalized for complex structures with multiple sublattices $i$ of $N_i$ sites and composition $x_i$. This entropic shortfall must be restored when calculating the free energy of an ensemble of finite size cells. We apply this correction for the bcc solid solution and the cI52 phases.

\section*{Results}
\subsection*{ Formation Enthalpy, \(\Delta H\)}

\begin{figure}[t]
  \centering
  \includegraphics[width=\linewidth]{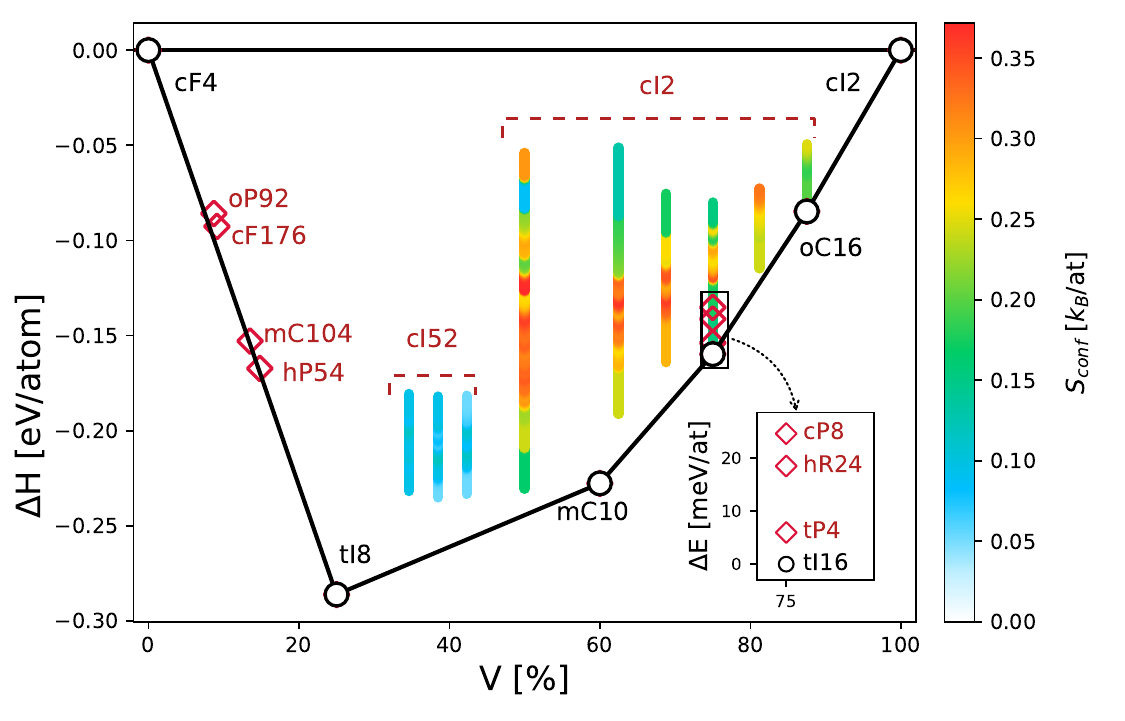}
  \caption{Enthalpies \(\Delta H_f^{(\tau)}\) evaluated at discrete compositions. Black circles on the convex hull represent ground states. Red diamonds are metastable line compounds. Columns in rainbow scheme represent the cubic \(\gamma\) brass structures in the homogeneity range centered on \phase{Al_8}{V_5} ($x=0.385$) and the bcc solid solution compositions ranging from $x=0.5$ to 1. At $x=0.75$, energy differences from the hull, \(\Delta E\) are shown in the inset. The colorbar represents configurational entropy \(S_{conf}(E)\), in units of \(k_B/at\), associated with cI2 and cI52 phases.  }
  \label{fig:deltaH}
\end{figure}

Figure~\ref{fig:deltaH} shows composition-dependent formation enthalpies determined from  DFT. Six ground states are identified: \phase{Al}{.cF4}, \phase{V}{.cI2} (both by definition), \phase{Al_3V}{.tI8}, \phase{Al_2V_3}{.mC10}, \phase{AlV_3}{.tI16}, \phase{Al_2V_{14}}{.oC16}.  Two of them, \phase{Al_2V_3}{.mC10} and \phase{Al_2V_{14}}{.oC16} were obtained using ATAT. \phase{AlV_3}{.tI16} was obtained by applying an infinitesimal lattice distortion to a dynamically unstable configuration of \phase{AlV_3}{.cI2}, followed by structural relaxation. The inset plot shows the competition among four structures at $x=0.75$. The nearest competitor to \phase{tI}{16} is \phase{tP}{4}, only $\Delta E = 5.95$ meV/at above the convex hull. Derived from the elastically unstable structure \phase{AlV_3}{.cP8}, \phase{AlV_3}{.hR24} is lower in formation enthalpy than its cP8 counterpart by 7.48 meV/at, but still lies \(\Delta E =\) 18.47 meV/at above the hull.

\phase{Al_{45}V_7}{.mC104} and \phase{Al_{23}V_4}{.hP54} lie only \(\Delta E = \) 1.19 and 2.27 meV/at above the convex hull, respectively. Another closely related pair, \phase{Al_{10}V}{.cF176} and \phase{Al_{10}V}{.oP92} ($x=0.091$ and $0.095$, respectively), differ in \(\Delta E\) by only 2.28 meV/at; the lower-energy \phase{Al_{10}V}{.cF176} remains \(\Delta E = 11.37 \) meV/at above the hull. \phase{Al_{10}V}{.oP92} is a symmetry-broken variant of cF184 with all guest sites occupied, while the cF176 structure is cF184 with all guest sites vacant.

For the disordered cI52 phase, we observe a common range in formation enthalpy with a spread of \(\sim 40\) meV, centered on $x=0.385$ (\phase{Al_8}{V_5}), bracketed by $x=0.346$ (\phase{Al_{17}}{V_9}) and $x=0.423$ (\phase{Al_{15}}{V_{11}}). As \(\Delta E \geq 32\) meV/at, cI52 remains unstable at low temperatures. The bcc solution exhibits a much larger enthalpy spread at lower V content because each of the 16 sites in the unit cell is partially occupied according to the nominal composition. At \phase{Al_8}{V_8} (the lowest Vanadium composition considered), the spread is $\sim 150$ meV, whereas at \phase{Al_2}{V_{14}} (excluding elemental Vanadium), it is only 14 meV. Correspondingly, these two compositions have 58 and 4 symmetry-distinct structures, respectively. With larger spreads in formation enthalpy, cI2 phases also have larger variations in configurational entropy \(S_{conf}\), shown in color contours (see Figure ~\ref{fig:deltaH}). For individual structures \(\tau\), \(S_{\text{conf}}^{(\tau)} = \ln \Omega^{(\tau)}\), where \(S_{conf}(E)\) is calculated by Gaussian smearing these discrete contributions with a width of 2 meV \cite{sarker2018high}.

\section*{Free Energy Comparison $T>0$}

\begin{figure}[t]
  \centering
  \includegraphics[width=\linewidth]{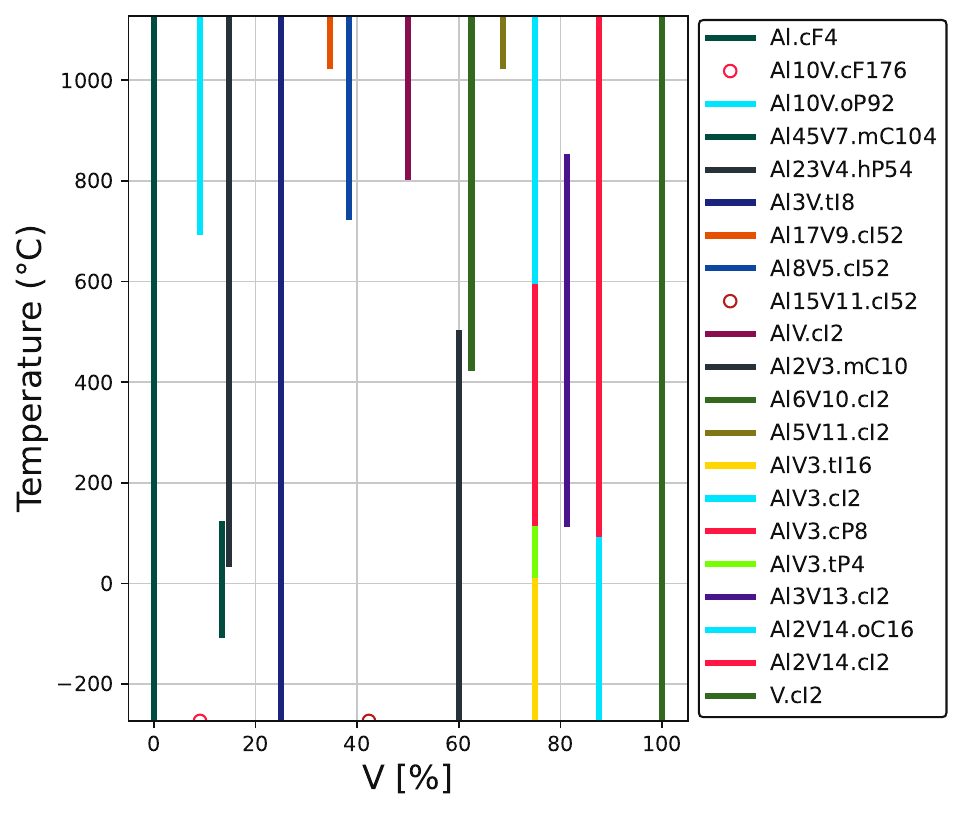}
  \caption{Phase stability at discrete compositions. Semi-circles at the bottom represent phases that are never stable at any temperature. The legend lists structures in order of increasing concentration of vanadium.} 
  \label{fig:sp}
\end{figure}

We first construct the convex hull of free energy at a discrete set of compositions to identify stable phases over temperatures 0-1373 K, corresponding to -273--1100 °C. Figure~\ref{fig:sp}  shows the stable temperature range at each composition. 
Consider the \phase{Al_{10}V}{.cF176} and \phase{Al_{10}V}{.oP92} structures. While neither lies on the hull at low temperatures, oP92 becomes stable at elevated temperatures owing to the vibrational entropy of its occupied cage atoms~\cite{albuhairan2025first}.
Next, consider \phase{Al_{23}V_4}{.hP54} and \phase{Al_{45}V_7}{.mC104}. Both phases have similar vibrational entropy, reflecting their similar composition and closely related structures.  mC104 becomes stable at lower temperatures than hP54, due to its slightly lower \(\Delta E\), that coexists with pure \phase{Al}{.cF4} and \phase{Al_3V}{.tI8}. Following a small $T$-region of coexistence between hP54 and mC104, as $T$ rises hP54 wins out owing to the combined high vibrational entropy of Al and low entropy of \phase{Al_3V}{.tI8}.
.

The large, negative enthalpy of \phase{Al_3V}{.tI8} gives it stability over the entire temperature range considered, and thereby insulates the Al-rich phases just discussed from the less Al-rich phases that we now examine. Between \phase{Al_3V}{.tI8} and \phase{Al_2V_3}{.mC10} lies the cI52 phase that we examine at three discrete compositions. The configurational entropy of partially occupied sites stabilizes this phase at elevated temperatures. \(\Delta E\) is smaller for cI52 compositions higher in the V content because of the slope of the tie-line joining tI8 to mC10. This helps \phase{Al_8V_5}{.cI52} gain stability first at \(T>700\)°C followed by the stability of \phase{Al_{17}}{V_9.cI52} at higher temperatures while \phase{Al_{15}V_{11}}{.cI52} remains marginally close to the convex hull.

\begin{figure}[t]
  \centering
  \includegraphics[width=\linewidth]{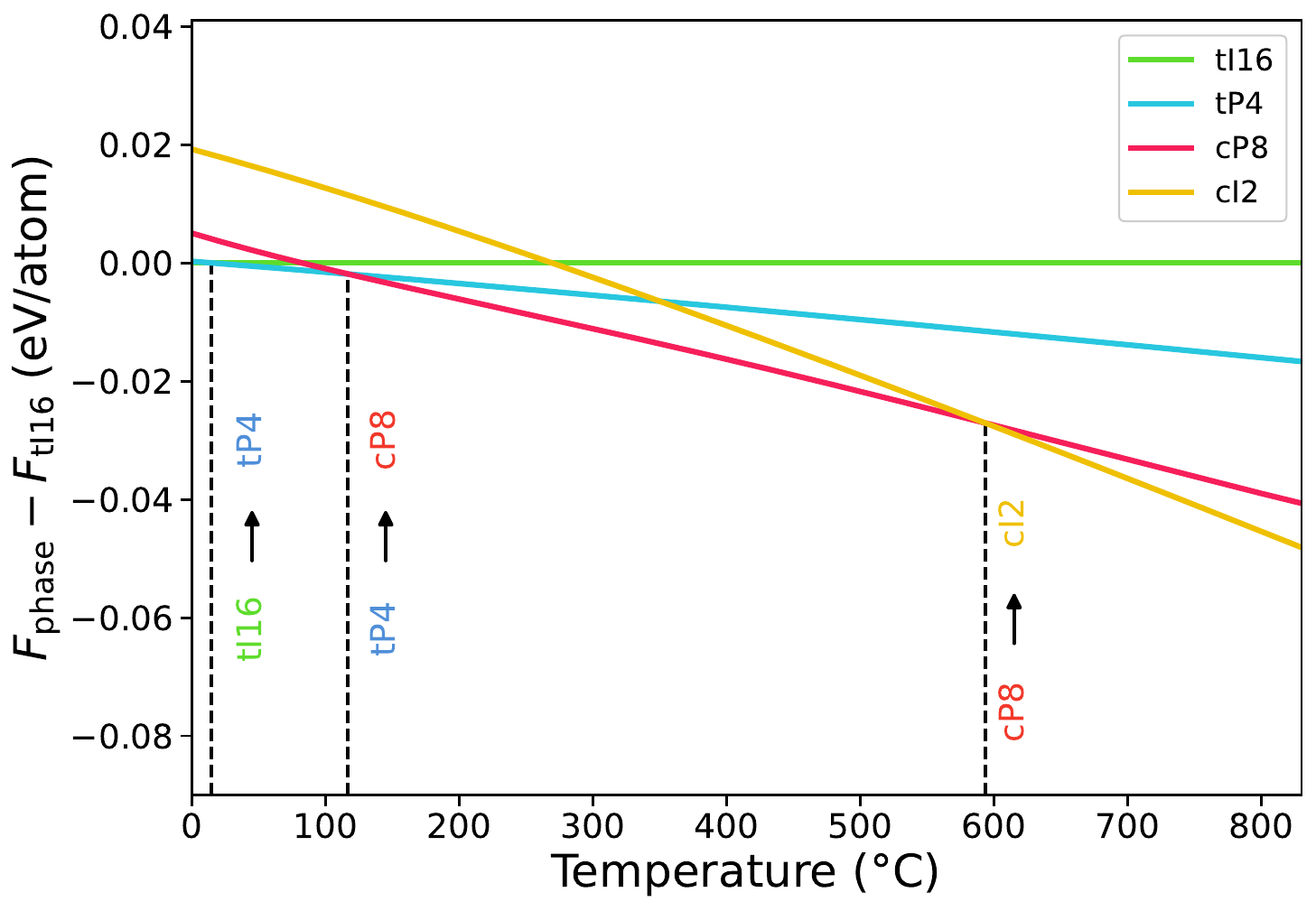}
  \caption{Free energy comparison between different crystalline phases and the disordered phase of \phase{Al}{V_3}. All the free energies in this plot are referenced to that of the tI16 phase. The transformations between competing phases are shown with dotted lines.} 
  \label{fig:alv3}
\end{figure}

In V-rich compositions, the newly proposed ground states \phase{Al_2V_3}{.mC10}, \phase{AlV_3}{.tI16}, and \phase{Al_2V_{14}}{.oC16} transition to the cI2 solid solution at elevated temperatures.  However, at \(x=0.75\), the transition is more complex (figure~\ref{fig:alv3}). First, the ground state tI16 transforms to tP4 at 16°C due to the higher vibrational entropy of the latter. A second transition follows at 114°C, where tP4 is replaced by cP8. cP8 itself is mechanically unstable at low temperatures: it transforms to hR24 under first-principles molecular dynamics below 100 K. Owing to its mechanical instability we applied the method of temperature-dependent effective potential ~\cite{Hellman2013,Knoop2024TDEP} in order to obtain its vibrational free energy. At high temperatures, the cI2 phase is stable due to its higher vibrational entropy, and the transition from cP8 to cI2 occurs at 585°C. Since we treat anharmonicity only for cP8, the transition temperatures for both tP4\(\to\)cP8 and cP8\(\to\)cI2 carry additional uncertainty, as indicated by the dashed lines in Figure~\ref{fig:alv3}.

\begin{figure}[htbp]
  \centering
  \includegraphics[width=0.83\linewidth, angle=-90]{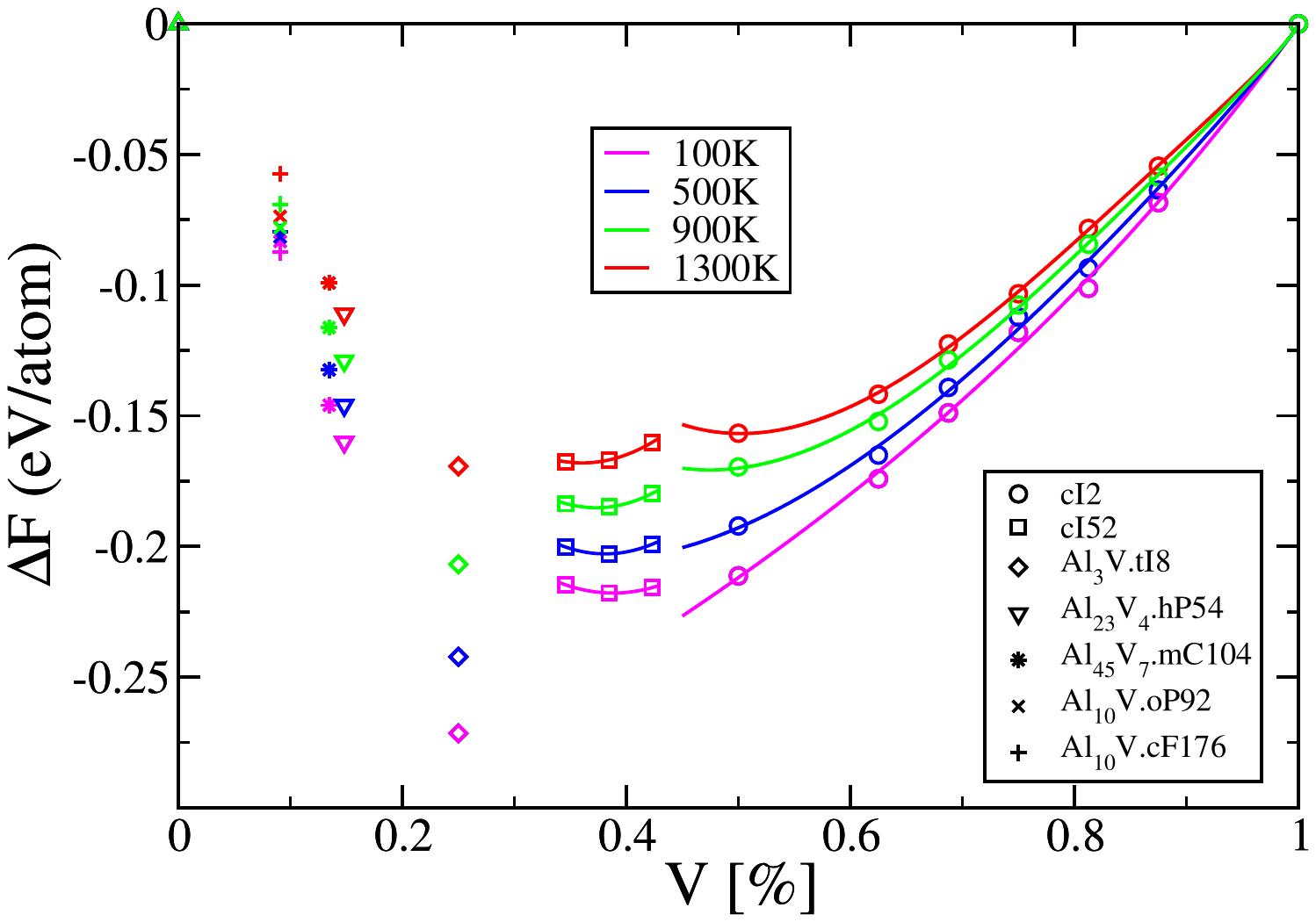}
  \caption{Free energy comparison between discrete structures (points) and continuous cI2 and cI52 models (lines). For clarity, free energies of V-rich line compounds have been omitted.} 
  \label{fig:Ffit}
\end{figure}

To determine a continuous phase boundary of the solid solution, we model its free energy as a continuous function of  $x$ and $T$. The discrete cI2 ensemble exhibits modest composition-dependent fluctuations in stability temperature (see Figure~\ref{fig:sp}), arising from the different enthalpic and entropic contributions of the configurations \(\{\tau\}\) sampled at each composition (see Fig.~\ref{fig:deltaH}). Such fluctuations should diminish in the thermodynamic limit.
 Our model combines the ideal-mixing entropy with an excess term $F_{\text{ex}}(x,T)$ that we take to be cubic in composition:
\begin{equation}
\begin{aligned}
F_{\mathrm{model}}(x,T)
&= k_{B}T\Big[x\ln x+(1-x)\ln(1-x)\Big] + F_{\text{ex}}(x,T),\\
F_{\text{ex}}(x,T)
&= a(T)\,x^{3} + b(T)\,x^{2} + c(T)\,x + d(T).
\end{aligned}
\end{equation}
Coefficients \(a, b, c\), and \(d\) were determined by least square fitting to $F_{\text{model}}$ for temperatures ranging from 0 K to 1373 K at intervals of 1 K . Figure~\ref{fig:Ffit} shows the model free energy function  from \(x=0.45\) to \(x=1\) for \(T=100,500,900,1300\) K. The model reproduces discrete data with residuals of \(\pm 2.5\)meV with an exception, \phase{AlV_3}{.cI2}, for which the residual is \(\pm 5\)meV. Similarly, we model the free energy function of the cI52 phase from \(x=0.345\) to \(x=0.425\).

\begin{figure}[htbp]
  \centering
  \includegraphics[width=0.95\linewidth]{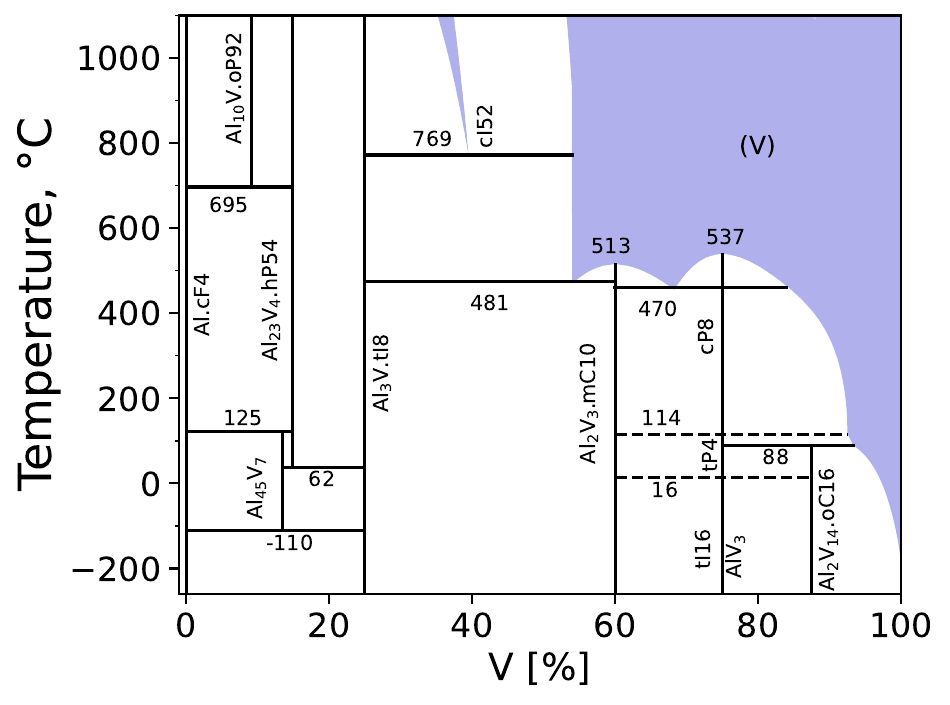}
  \caption{Modeled Phase diagram (excluding liquid phase). Dashed lines indicate uncertainty in the transition temperatures of \phase{Al}{V_3} phases.} 
  \label{fig:PB}
\end{figure}

To evaluate the global phase diagram over the full range of composition and temperature, we evaluate the free energy of each phase at a given temperature and identify those states that lie in the convex hull. The result is shown in  \textbf{Figure}~\ref{fig:PB}.

\section*{Discussion}

In this paper we calculate the temperature-dependent free energies of known and hypothetical phases in the \phase{Al_{1-x}}{V_x} binary alloy system. Special attention is paid to the phases \phase{Al_8}{V_5}.cI52 and the V-rich BCC solid solution that exhibit sites of mixed (Al,V) occupation. In these cases we apply the method of full enumeration that determines all possible arrangements of species within a cell of a given size, and sums the full partition function weighted by structural multiplicity, enthalpy, and vibrational and electronic free energies. However, the finite size artificially lowers the entropy of mixing, and we introduce a correction for this effect. We then calculate the composition- and temperature-dependent phase diagram from the convex hull of the free energies.


The Al-rich and V-rich regions of \phase{Al_{1-x}}{V_x} differ strongly in character. The Al-rich side contains several line compounds with narrow composition ranges and complex crystal structures, often including icosahedral motifs. We model these as line compounds. Because we apply only harmonic phonon modeling and neglect thermal expansion, our present results differ from experiment more strongly than in our prior paper~\cite{albuhairan2025first}. This is particularly true in the case of \phase{Al_{10}}{V} where we only address the cases of fully vacant and fully occupied cages, and the minimum temperature of stability is 500°C higher than previously found \cite{albuhairan2025first}. Indeed, we mainly include these Al-rich phases for completeness, as the V-rich side is our primary interest.

The V-rich BCC solid solution exists over a broad range of composition. At low temperature the composition range must transform into a series of one or more ordered states. We found three such states. One of them, \phase{Al_2}{V_3}.mC10, which was identified with the aid of ATAT~\cite{van2002alloy}, has the composition of 60\% vanadium that coincides with the limiting solid solution range shown in Fig.~\ref{fig:AlVPD}. Notably, this phase exhibits a deep and narrow pseudogap at the Fermi energy, suggesting an electronic structure explanation for this limit. The ground state at \(x=0.75\), \phase{Al}{V_3}.tI16, and the intermediate phase \phase{Al}{V_3}.tP4 arose in our full enumeration of configurations based on a 16-atom supercell of the BCC lattice. The lattice parameters of neither of them correspond to these of the low T-P tetragonal \phase{Al}{V_3} reported by \cite{leger1973pressure}. The experimentally reported \(a/c\) is 0.645, but the corresponding values for tI16 and tP4 are 0.895 and 0.5, respectively. Thus, the identity of the experimentally reported tetragonal phase remains an outstanding issue for the phase diagram.

Based on full enumeration of chemical ordering on a 26-atom primitive cell, no configuration of \phase{Al_8V_5}{.cI52} is part of 0 K convex hull. In fact, there is no low temperature stable structure at any of the cI52 compositions. The experimental work \cite{richter2000v} finds the cI52 phase stable at 1050°C for \(x=0.363,0.385,\) and \(0.395\). Further, \cite{carlson1954aluminum} shows that the stable cI52 region moves toward the Al-richer side and increases in width at higher temperatures. Our modeled phase boundary of cI52 in Figure~\ref{fig:PB} is consistent with both experimental conclusions.

\begin{figure}[htbp]
  \centering
  \includegraphics[width=0.8\linewidth, angle=-90]{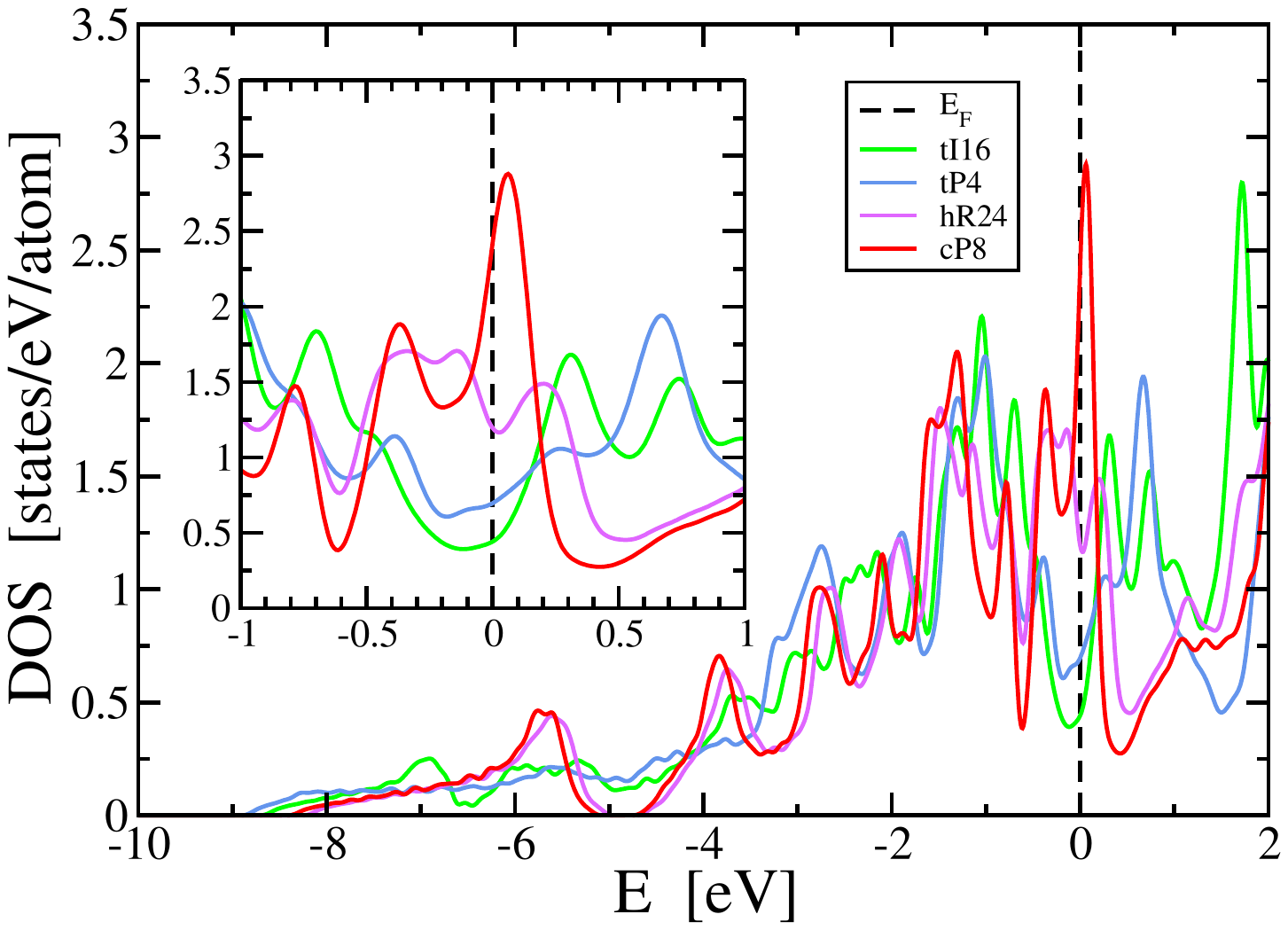}
  \caption{Comparison in DOE (density of states) between different crystalline phases of \phase{Al}{V_3}. The inset shows the location of the pseudo-gap relative to the Fermi level for each structure.} 
  \label{fig:alv3-dos}
\end{figure}

\phase{Al}{V_3}.cP8 is an A15 structure whose existence was described as uncertain~\cite{murray1989}. Fig.~\ref{fig:AlVPD} gives it a wide composition range, but this should not occur, since Al and V each occupy distinct sites of coordination number 14 and 12, respectively, so we model it as a line compound. However, it is predicted to be mechanically unstable and transforms at low temperatures into a structure of Pearson type hR24 creating a pseudogap at the Fermi level in its electronic density of states (see inset of Figure~\ref{fig:alv3-dos}). In fact, all line compounds  at \(x=0.75\) exhibit a pseudogap-like depletion near \(E_F\). With decreasing temperature, \phase{Al}{V_3} favors the position of this pseudogap further below the Fermi level, corresponding to the transitions from cP8 to tP4 to tI16 at low temperatures.

In conclusion, we have examined the composition vs. temperature phase diagram of the Al-V binary alloy system. Through density functional theory-based total energy and statistical mechanical calculations we argue that phases \phase{Al_{10}}{V}, \phase{Al_{45}{V_7}}, \phase{Al_{23}{V_4}}, and \phase{Al_8}{V_5} are not stable at low temperatures but are entropically stabilized at elevated temperatures. We establish that the V-rich BCC solid solution is stable with a broad composition range but only at high temperatures. At low temperatures it transforms into a series of ordered phases whose structures we identify. The A15 compound \phase{Al}{V_3}.cP8 is found to be mechanically unstable at low temperatures but is stabilized by its anharmonic vibrational entropy over an intermediate temperature range.

\appendix
\section{Appendix: Structures Considered}
\label{app:structures}

VASP calculations for \phase{Al_{1-\text{x}}}{V_{\text{x}}} were performed for the following structures. We give the chemical formulas, lattice constants, Pearson types, space group numbers, and Wyckoff sites. VASP format POSCAR files and cif files for all structures are available at \url{https://alloy.phys.cmu.edu}.

\subsection*{Ground State Structures}

\small
\setlength{\tabcolsep}{4pt}
\renewcommand{\arraystretch}{1.1}

\centering
\begin{tabular}{||lllll||}
\toprule
Composition & Geometry & \makecell[l]{Lattice param \\}  & Pearson Type & Group No. \\
\midrule
\(x=0\): Al & cubic & \(a=4.039\) & cF4   & 225 \\ \midrule
\(x=0.25\): Al$_3$V & tetragonal & \makecell[l]{\(a=3.765\)\\\(b=8.307\)} & tI8  & 139 \\ \midrule
\(x=0.6\): Al$_2$V$_3$ & monoclinic & \makecell[l]{ \(a=9.290\) \\\(b=3.206\) \\\(c=6.566\) \\ \(\alpha = \gamma = 90\) \\ \(\beta = 133.511 \)}  & mC10 & 12  \\ \midrule
\(x=0.75\): AlV$_3$ & tetragonal & \makecell[l]{\(a=5.825\) \\\(b=6.506\) } & tI16 & 139  \\ \midrule
\(x=0.875\): Al$_2$V$_{14}$ & orthorhombic & \makecell[l]{ \(a=11.924\) \\\(b=6.022\) \\\(c=3.047\)} & oC16 &  65 \\ \midrule
\(x=1\): V & cubic & \(a=2.994\)  & cI2 &  229 \\
\bottomrule
\end{tabular}

\subsection*{Other Line Compounds}

\begin{tabular}{||lllll||}
\toprule
Composition &  Geometry  & Lattice Constant & Pearson Type & Group No. \\
\midrule
\(x=1/11\): Al$_{10}$V & cubic & \(a=14.449\) & cF176 & 227    \\ \midrule
\(x=1/11\): Al$_{10}$V & orthorhombic & \makecell[l]{\(a=10.231\) \\\(b=10.242\) \\\(c=14.472\)} & oP92 &  62  \\ \midrule
\(x=4/27\): Al$_{23}$V$_4$ & hexagonal & \makecell[l]{\(a=7.676\) \\\(b=17.025\) \\ \(\alpha = \beta = 90\) \\ \(\gamma = 120\)} & hP54  & 194  \\ \midrule
\(x=7/52\): Al$_{45}$V$_7$ & monoclinic & \makecell[l]{ \(a=20.850\)  \\\(b=7.606\) \\ \(c=11.077\) \\ \(\alpha = \gamma = 90\) \\ \(\beta = 106.964 \)} & mC104 & 12 \\ \midrule
\(x=0.75\): AlV$_3$  & tetragonal & \makecell[l]{\(a=3.078\) \\\(b=5.880\)} & tP4 & 123 \\ \midrule
\(x = 0.75\): Al$V_3$ & cubic & \makecell[l]{\(a=4.803\)} & cP8 & 223 \\ \midrule
\(x=0.75\): AlV$_3$ & hexagonal & \makecell[l]{\(a=b=6.947\) \\ \(c=7.952\) \\ \(\alpha =\beta = 90\) \\ \(\gamma = 120\)} & hR24 & 167 \\
\bottomrule
\end{tabular}

\subsection*{Wyckoff Sites of all Structures}

\begin{flushleft}

\textbf{Al.cF4}: \textbf{Al}: 4a (0,0,0).

\(\mathbf{Al_3V}\)\textbf{.tI8}: \textbf{Al:} 2b (0,0,0.5) and 4d (0,0.5,0.25). \textbf{V:} 2a (0,0,0).

\(\mathbf{Al_2V_3}\)\textbf{.mC10}: \textbf{Al:} 4i (0.191,0,0.782). \textbf{V:} 2a (0,0,0) and 4i (0.606,0,0.378).

 \(\mathbf{AlV_3}\)\textbf{.tI16}: \textbf{Al:} 4e (0,0,0.209). \textbf{V:} 4d (0,0.5,0.25) and 8h (0.726,0.726,0).

\(\mathbf{Al_2V_{14}}\)\textbf{.oC16}: \textbf{Al:} 2b (0.5,0,0).\textbf{V:} 2a (0,0,0), 8q (0.869,0.231,0.5), and 4g (0.257,0,0).

\(\mathbf{Al_{10}V}\)\textbf{.cF176}: \textbf{Al:} 48f (-0.0152,0.125,0.125), 96g (0.56,0.56,0.823), and 16d (0.5,0.5,0.5). \textbf{V:} 16c (0,0,0).

\(\mathbf{Al_{10}V}\)\textbf{.oP92}: \textbf{Al:} 8d (0.86,0.39,0.375), 8d (0.36,0.890,0.875), 4c (0.500,0.25,0.265), 4c (0.500,0.25,0.015), 4c (0.528,0.25,0.632), 8d (0.7653,0.384,0.56), 8d (0.633,0.515,0.690), 8d (0.132,0.013,0.190), 8d (0.262,0.882,0.059), 4c (0.87,0.25,0.072), 8d (0.001,0.119,0.677), 4c (0.631,0.25,0.427), 4c (0.761,0.25,0.756), 4a (0.0,0.0,0.0). \textbf{V:} 4c (0.750,0.25,0.25), 4b (0,0,0.5).

\textbf{Al\(_{23}\)V\(_4\).hP54}: \textbf{Al:} 12k (0.875,0.75,0.884), 6h (0.874,0.748,0.25), 12k (0.790, 0.579,0.471), 4f (0.333,0.667,0.384), 12k (0.541,0.082,0.334). \textbf{V:} 2a (0,0,0), 6h (0.219,0.438,0.25).

\textbf{Al\(_{45}\)V\(_7\).mC104}: \textbf{Al:} 2d (0,0.5,0.5), 4i (0.521,0,0.757), 8j (0.536,0.313,0.205), 8j (0.545,0.306,0.623), 8j (0.569,0.182,0.0103), 4i (0.082,0.0,0.441), 4i (0.09,0.0,0.845), 4i (0.615,0.0,0.228), 4i (0.625,0.0,0.638), 4i (0.129,0.0,0.113), 8j (0.165,0.306,0.856), 8j (0.180,0.185,0.351), 8j (0.190,0.185,0.621), 8j (0.706,0.185,0.133), 4i (0.707,0.0,0.490), 4i (0.224,0.0,-0.008). \textbf{V:} 2a (0,0,0), 8j (0.585,0.170,0.419), 4i (0.749,0,0.761).

\(\mathbf{AlV_3}\)\textbf{.tP4}: \textbf{Al:}  1a (0,0,0). \textbf{V:}  2h (0.5,0.5,0.273), 1b (0,0,0.5).

\(\mathbf{AlV_3}\)\textbf{.cP8}: \textbf{Al:}  2a  (0,0,0). \textbf{V:}  6d (0.25,0.5,0).

\(\mathbf{AlV_3}\)\textbf{.hR24}: \textbf{Al:}  6b  (0,0,0). \textbf{V:}  18e (0.737,0,0.25).

\end{flushleft}

\bibliographystyle{apsrev4-2}
\bibliography{references}
\end{document}